\documentclass[a4paper]{spie}  
\usepackage[]{graphicx,subfigure,float, amssymb,color}

\newcommand{\eqref}[1]{Eq.~(\ref{#1})}

\newcommand{\figref}[1]{Fig.~(\ref{#1})}
\newcommand{\tabref}[1]{Tab.~(\ref{#1})}

\title{Development of TES-based detectors array for the X-ray Integral Field Unit (X-IFU) on the future X-Ray Observatory Athena.} 

\author{L. Gottardi\supit{a}, H. Akamatsu\supit{a}, D. Barret\supit{b}, M.P. Bruijn\supit{a}, R.H. den Hartog\supit{a}, J-W. den Herder\supit{a}, H.F.C. Hoevers\supit{a},  M. Kiviranta\supit{c}, J. van der Kuur\supit{a}, A.J. van der Linden\supit{a}, B.D. Jackson\supit{a}, M. Jambunathan\supit{a}, M.L. Ridder\supit{a} 
\skiplinehalf
\supit{a} SRON National Institute for Space Research, \\ Sorbonnelaan 2, 3584 CA Utrecht, The Netherlands \\
\supit{b} Institut de Recherche en Astrophysique et Planétologie (IRAP), \\ Avenue du Colonel Roche.BP 44346, 31028 Toulouse Cedex 4, France\\
\supit{c} VTT\\ Tietotie 3, 02150 Espoo, Finland\\
}

\begin{document} 
  \maketitle 

\begin{abstract}
We are developing transition-edge sensor (TES)-based microcalorimeters for the X-ray Integral Field Unit (X-IFU) of the future 
European X-Ray Observatory {\it Athena}. 
 The microcalorimeters are based on TiAu TESs coupled to $250\mu\mathrm{m}$ squared, AuBi absorbers. We designed and fabricated devices with different contact 
geometries between the absorber and the TES to optimise the detector performance and with different wiring topology to mitigate the self-magnetic 
field. The design is tailored to optimise the performance under Frequency Domain Multiplexing. 
In this paper we review the main  design feature of the pixels array and we report on the performance of the  18 channels, 2-5MHz frequency domain multiplexer that will be used to characterised the detector array.  

\end{abstract}



\section{INTRODUCTION}
\label{sec:intro}  

Transition-edge sensor (TES)-based microcalorimeters are the chosen technology for the detectors array of the X-ray Integral Field Unit (X-IFU) on board of the future European X-Ray
 Observatory {\it Athena}. The X-IFU is a 2-D imaging integral-field spectrometer operating in the soft X-ray band ($0.3-12\,\mathrm{keV}$).
The detector consists of an array of 3840 TESs coupled to X-ray absorbers and read out using Frequency Domain Multiplexing (FDM) in the MHz bandwidth. 
The {\it Athena} proposed design calls  for devices  with  an high filling-factor, high quantum-efficiency, high count-rate capability  and an energy resolution of $2.5\, \mathrm{eV}$ at $5.9\, \mathrm{keV}$. 

The requirements for the most relevant pixel-related parameters for the {\it Athena} X-IFU instrument are listed in \tabref{xifutable} 

\begin{table}[htbp]
  \begin{center} \small
    \begin{tabular}{l l}  
\hline 
\vspace{0.06 cm} {\bf Parameters} & {\bf Requirements}  \\
\hline
\vspace{0.1 cm} Energy range & $0.2-12\,\mathrm{keV}$\\
\vspace{0.1 cm}   Energy resolution: $E < 7\, \mathrm{keV}$ & $2.5\, \mathrm{eV}$ \\
\vspace{0.1 cm}  Pixel size &  $250 \times 250\, \mu \mathrm{m}^2$)\\
\vspace{0.1 cm} Field of view & $5^\prime$ (diameter) (3840 pixels)\\
\vspace{0.1 cm} Quantum efficiency @ $6\, \mathrm{keV}$ & $>90\%$\\
\vspace{0.1 cm} Count rate capability - faint source & 1 mCrab ($>80\%$ high-resolution events)\\
\vspace{0.1 cm} Count rate capability - bright source & 1 Crab ($>30\%$ low-resolution events)\\
\hline\\
     \end{tabular}
     \caption{Key performance requirements for the {\it Athena} X-IFU pixels array}
     \label{xifutable}
  \end{center}
\end{table}

\begin{figure}[htbp]
    \centering
    \includegraphics[height=6.cm]{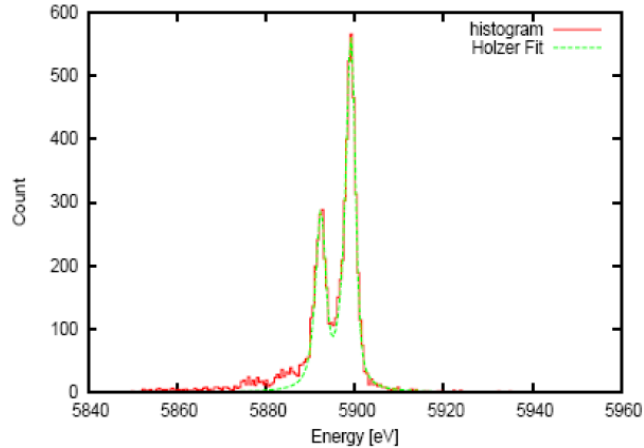}
    \caption[example] 
   { \label{fig:K-alphaCuBi} $^{55}Fe$ spectrum taken with a $100 \times 100\, \mu\mathrm{m}^2$, $1\,\mu\mathrm{m-thick}$ Cu and  $2.6\,\mu\mathrm{m-thick}$ Bi absorber.}
   \end{figure}

At SRON we  develop microcalorimeters arrays based on superconducting TiAu bilayers TES deposited on suspended SiN membranes thermally  coupled to Cu/Bi film absorbers. The Bismuth layer is added on top of the Cu layer to achieve sufficient stopping power without increasing significantly the total absorber heat capacity. In the past we have shown that pixels made of TiAu TESs coupled to a $1\,\mu\mathrm{m-}$thick Cu and $2.6\,\mu\mathrm{m-}$thick Bi $100\times100\,\mu \mathrm{m}^2$ central absorber could achieve high energy resolution \cite{Dirks09} when DC biased. The K-alpha spectrum of a Fe-55 radioactive source detected with this pixel is shown in \figref{fig:K-alphaCuBi}. We measured an  energy resolution of $2.5\,\mathrm{eV}$ at $5.9\,\mathrm{keV}$ under DC bias. 

\begin{figure}[htbp]
    \centering
    \includegraphics[height=6cm]{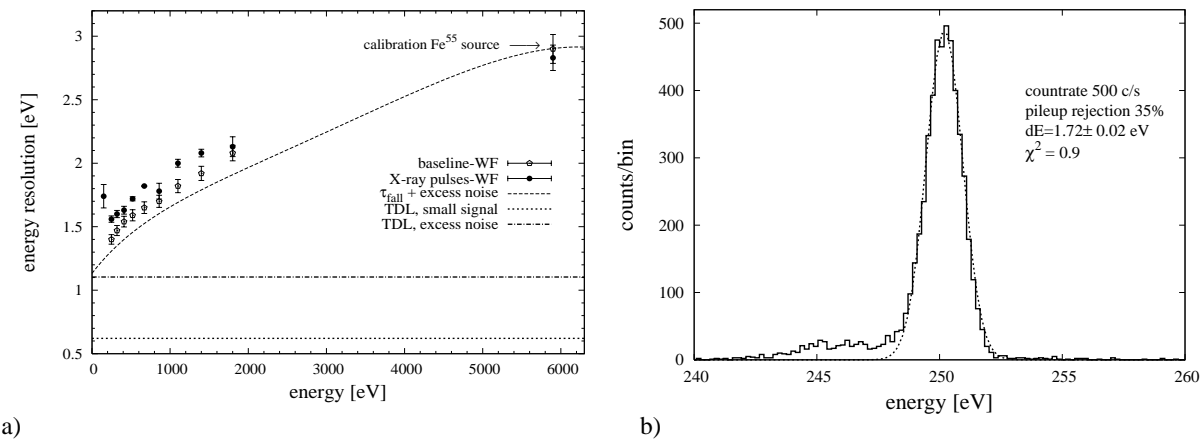}
    \caption[example] 
   { \label{fig:BESSYpixel} ({\bf a.}) Energy resolution as a function of incident photon energy for a pixel with $100 \times 100 \mu\mathrm{m}^2$, $1\mu$-thick Cu absorber. ({\bf b.}) Energy spectrum at $250\, \mathrm{eV}$ for a photon flux of about $500\, \mathrm{photons}/\mathrm{sec}$. }
   \end{figure}

An identical  pixel with central Cu absorber has been extensively characterised in a syncrotron facility with X-ray photons with energy ranging from 150 to $2000\, \mathrm{eV}$ \cite{Gottardi08,Takei08}. We demonstrated an high X-ray energy resolution and high count-rate capability as shown in \figref{fig:BESSYpixel}, where the energy resolution as a function of the incident photon as well as the energy spectrum for a photon flux of about $500\, \mathrm{photons}/\mathrm{sec}$ are plotted. 
At 250 eV and low flux rate  we measured the best energy resolution of $1.52\pm 0.02\,\mathrm{eV}$, which only slightly degraded to $1.78\pm 0.02\, \mathrm{eV}$  with  $500\, \mathrm{photons}/\mathrm{sec}$ hitting the detector \cite{Gottardi08}.
This pixel has been fully characterised under AC bias as well and we reported a best energy resolution of 3.7eV \cite{Gottardi09}. 
The performance of these pixels is however limited by  a poor filling factor (about $20\%$), and a relatively low absorption efficiency, where a maximum of $74\%$ at $5.9\,\mathrm{keV}$ with a $2.6\,\mu\mathrm{m}$-thick Bi layer has been achieved.
Moreover they have never been tested so far with a Frequency Domain Multiplexer operating at MHz frequencies.   
The most representative results of a MHz Frequency Domain Multiplexing (FDM) read-out of X-IFU-like TES microcalorimeters  is reviewed by Ravera {\it et al}\cite{Ravera14}, where it is shown that an energy resolution of $3.6\,\mathrm{eV}$ at $5.9\,\mathrm{keV}$ has been demonstrated\cite{Akamatsu14} using detectors fabricated at Goddard \cite{Iyomoto08}.

We have recently started a new development program aiming at the fabrication of large array detectors, which  will meet the X-IFU instrument requirements when operating with the MHz-FDM read-out.
In the following session we present the current status of the fabrication of arrays with large CuBi and AuBi absorber and show the performance of the FDM set-up that will be used for the single pixel  characterisation.

\section{MICROCALORIMETER ARRAY FABRICATION} 
\label{sec:litho}
  Here we highlight the improvements made in the process fabrication of the micro-calorimeter array. The renewed mask design  consists of devices focusing on absorber coupling, alpha tuning, coplanar/strip line wiring configuration and self-magnetic field mitigation wiring topology.

\begin{figure}[htbp]
\centering
\includegraphics[width=1.0\textwidth,keepaspectratio,
  angle=0]{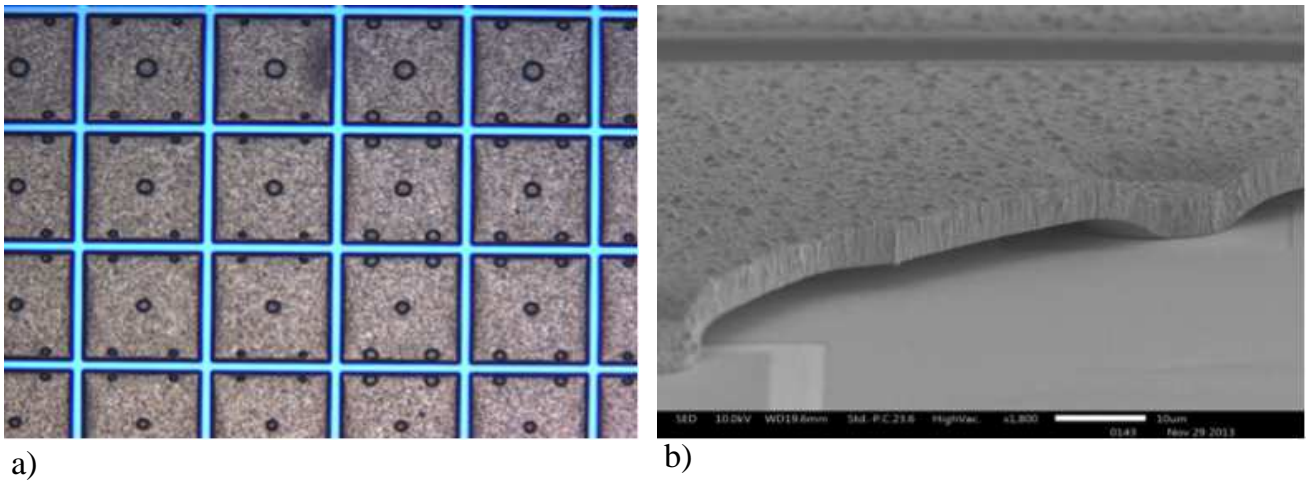}
\caption{ ({\bf a.}) Optical image of a 40x40  pixel array with  $240 \times 240\, \mu \mathrm{m}^2$ Bi absorbers.({\bf b.}) SEM image of a mushroom shaped Bi absorber hat  connected to the Au TES at the stem regions. \label{BiABS}}
\end{figure}

Earlier experiments showed that, by optimising the image reversal resist, Bi deposition and liftoff procedure, a free hanging mushroom absorber with only one stem attached to the TES for the thermal contact and 4 supporting ones   to the substrate can be fabricated. \figref{BiABS} shows optical images of test pixels from a $40\times 40$ array where the absorber is supported from the substrate by 4 stems with a diameter size of $5\mu\, \mathrm{m}$.

\begin{figure}[htbp]
\centering
\includegraphics[width=1.0\textwidth,keepaspectratio,
  angle=0]{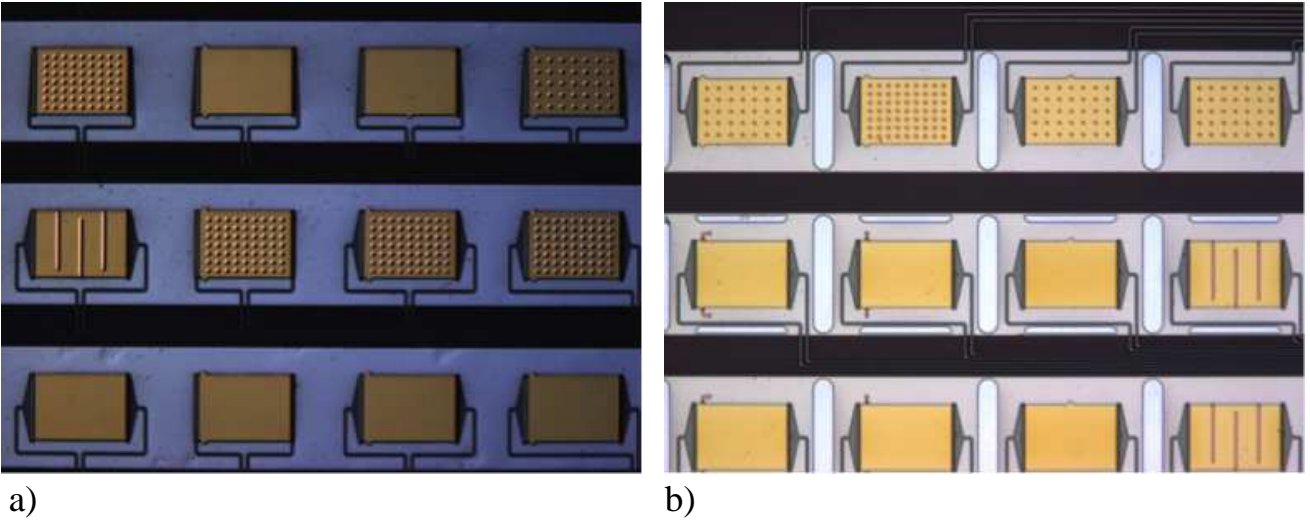}
\caption{ ({\bf a.}) Optical image of micro-calorimeter $5\times5$ array 
with TES, wiring  and Cu dots on a SiN membrane. ({\bf b.}) Optical image of the SiN slots on the micro-calorimeter array. \label{TESfoto}}
\end{figure}

In order to provide a weak thermal link to the cooling bath, the calorimeter must be fabricated on the silicon nitride membrane unlike solid silicon substrate.
Recently, we have fabricated the TES and the wiring layer on the SiN membrane. SiN slots and Cu dots fabrication( for alpha tuning) is also implemented as shown in \figref{TESfoto}.

The current work will be finalised by fabricating the mushroom shaped Cu-Bi absorber hat with foot region on the SiN membrane to demonstrate the absorber coupling.
The future area of focus will be in developing  the electroplated Au process to replace the Cu in the absorber for a further optimisation of the pixel performance  according to the X-IFU requirements.

\section{EXPERIMENTAL SET-UP} 
\label{sec:setup}

An FDM multiplexer similar to the one described by Gottardi {\it et al.}\cite{Gottardi12bolo} is
mounted on a removable probe of a helium-free dilution  refrigerator from Leiden  Cryogenics. It consists of a low magnetic impurity copper bracket
mounting on top  a two-stage SQUID amplifier, the LC filters and the TES array chips and a printed
circuit board for the electrical connection. One thermometer and
one heater are glued at the bottom of the copper plate for the
read-out and stabilisation of the temperature. The FDM set-up can be used both with ultra-low noise equivalent power (NEP) TES bolometers  and with high energy
resolving power X-ray microcalorimeters. The former requires very low
parasitic power loading ($<1\,\mathrm{fW}$), which is achieved by means of light blocking
filters in the signal loom feedthrough and a  light-tight
assembly. The X-ray microcalorimeters are very sensitive to magnetic
fields and their performance is optimal at static field lower than  
$1\,\mu \mathrm{Tesla}$).
 Special care has been taken to design the magnetic
shielding and to improve the uniformity of  the applied magnetic field
across the array. The TES arrays chip fits into a superconducting Helmholtz coil fixed at one end of the bracket. The coil is used to generate an
uniform perpendicular magnetic field over the whole pixels array. 
 The circuitry PC-board currently allows to read-out  18 pixels under ac 
 bias in a FDM configuration.
The SQUID chip is fixed to the
copper bracket using  high thermal conductivity glue (Ge Varnish). 
The TES array chip is connected via a superconducting interconnection
chip to the lithographic high-Q
 LC resonators arrays developed at SRON \cite{Bruijn12}. The nominal inductance of
 the coil used in  each filter is $L=400\,\mathrm{nH}$, while the
 capacitance's C are designed such that the  frequencies
 $f_0=\frac{1}{2\pi\sqrt{LC}}$ are spread at a constant interval of
 about $200\,\mathrm{kHz}$ in the range from 2 to $5\,\mathrm{MHz}$. The main purpose of the FDM test facility described here is to increase the experimental throughput for single pixel characterization. The FDM demonstrator for the {\it Athena} X-IFU instrument is described elsewhere \cite{denHartog14} .


The shielding of the external magnetic field is achieved by inserting the probe hosting the FDM set-up into a double-shield configuration made of cryoperm and lead.  can wrapped by few layers of metallic glass  tape. 

For the FDM read-out we use a low noise two-stage SQUID array provided by VTT. 
We use a VTT $180\times 4$ SQUIDs array  as {\it amplifier} SQUID and a low input inductance VTT 6-subloop fractional-turn SQUID as the {\it front-end} SQUID connected to the LC resonators and the TES array. \cite{Kiviranta13}. The {\it front-end} SQUID is voltage bias by applying a DC bias current to a $1\,\Omega$ shunt resistor connected in series between
the lower-SQUID output to the upper-SQUID input coil.
Both SQUID arrays can be  operated in the standard
not-linearized mode or in the current sensing mode  with
on chip linearization \cite{Kiviranta08}.
The SQUID chips are thermally coupled to the bracket by means of several Au bonding. 

The SQUIDs array signal is  read out using low-noise room temperature FEE
developed at SRON operating both in open-loop mode or in Baseband Feedback mode \cite{Hartog12}.  

\section{FDM READ-OUT CHARACTERISATION} \label{sec:fdmreadout}

Here we present the results of the two-stage SQUID characterisation at
$T=25\,\mathrm{mK}$ without LC resonators and TES microcalorimeters  connected to the {\it front-end} SQUID
input coil. 

\begin{figure}[htbp]
\centering
\includegraphics[width=1.\textwidth,keepaspectratio,
  angle=0]{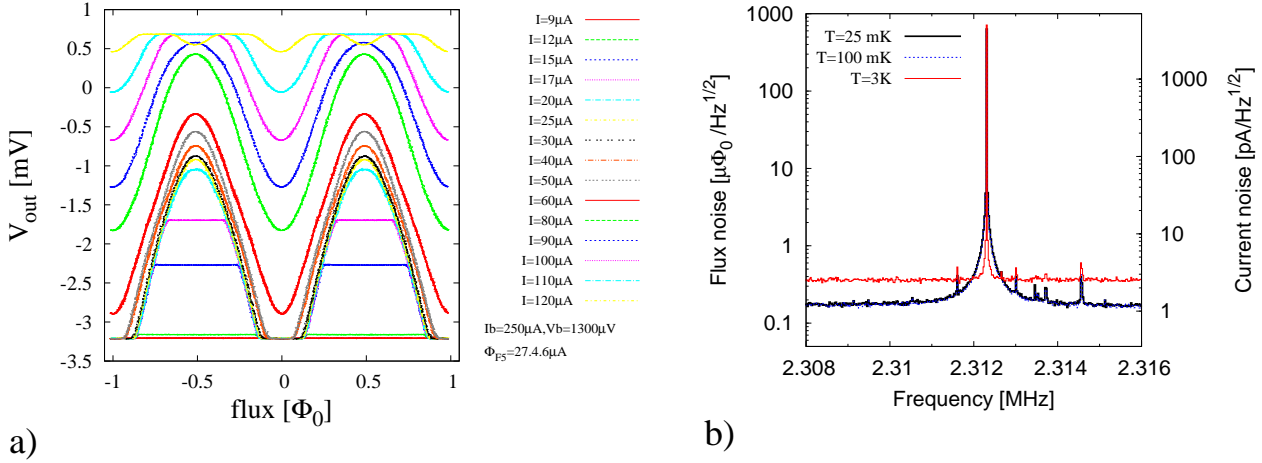}
\caption{ ({\bf a.}) Voltage-to-flux characteristics of the two-stage SQUIDs configuration operating at 20mK. ({\bf b.}) Flux and current noise spectra density for the VTT two-stage SQUIDs  measured at at $3\,\mathrm{K}$ and $20\, \mathrm{mK}$ with a calibration signal at $2.3\,\mathrm{MHz}$.  \label{VTTvphinoise}}
\end{figure}

In \figref{VTTvphinoise} we show the voltage-to-flux characteristics of the
two-stage SQUIDs measured as a function of the bias current of the
{\it front-end} SQUID. The two-stage SQUIDs array shows clean
V-$\Phi$ curves as well with a maximum voltage modulation of $2.4\,\mathrm{mV}$ at
a lower-SQUID bias current of $50\,\mu\mathrm{A}$. 

We measured the current noise of the two-stage SQUID at three
different temperature, respectively $3\,\mathrm{K}$,$100\,\mathrm{mK}$ and $20\,\mathrm{mK}$. 
Two calibration tones at $12\,\mathrm{kHz}$ and $2\,\mathrm{MHz}$  have
been applied via the feedback coil to monitor the SQUID gain as a
function of temperature. The magnitude of the calibration tone
correspond to an applied flux to the lower SQUID of $0.01\,\Phi_0$.
In \figref{VTTvphinoise}.a the spectral flux and current density taken
at temperature of $3\,\mathrm{K}$ and $25\,\mathrm{mK}$ is shown.The current noise has been
calculate using the nominal input sensitivity of the front-end SQUID array equal to
$M^{-1}=7\,\mu \mathrm{A}/\Phi_0$ 
In \figref{VTTvphinoise}.b a zoom-in of the SQUID noise spectra around
the $2\,\mathrm{MHz}$  calibration tone is shown.

We measured a flux noise equal to $3.5\cdot10^{-7}\,\Phi_0/Hz^{1/2}$ and
$1.8\cdot10^{-7}\,\Phi_0/Hz^{1/2}$ at $3\,\mathrm{K}$ and $20\,\mathrm{mK}$ respectively. This
corresponds to an energy resolution of
$\epsilon=S_{\Phi}/2L_{sq}\simeq9\,\hbar$, where an $L_{sq}=70\,\mathrm{pH}$ is assumed.
We believe the measured noise is limited by the room temperature
amplifier and the read-out acquisition system. A relatively straight
forward optimisation of the room temperature set-up should
lead to a further improvement of the SQUID sensitivity.  
The quantum limited flux sensitivity for the {\it front-end} SQUID array is 
$\sqrt{S_{\Phi,QL}}=6\cdot10^{-8}\,\Phi_0/Hz^{1/2}$, which should be achievable if the
temperature of the SQUID shunt resistors can be cooled at temperature
below $150\,\mathrm{mK}$.   
The measured current noise at the SQUID input is
$1.2\cdot10^{-12}\,A/Hz^{1/2}$ at $20\,\mathrm{mK}$, which corresponds to a coupled
energy resolution of $\epsilon_{ii}=\frac{L_{in}}{2}S_{ii}\sim
20\,\hbar$, where $L_{in}\sim 3\,\mathrm{nH}$ is the input inductance of
the SQUID. 
The two-stage SQUID was then tested in combination with 18 high-$Q$ $LC$ resonators and an array of TES-bolometers.
We measured a SQUID input current noise of $3\,\mathrm{pA}/\sqrt{Hz}$. The contribution of the SQUID noise on the energy resolution for a single pixel read-out has been estimated to be better than $0.1\,\mathrm{eV}$. The combination of low input inductance and low input  current noise of the {\it front-end} SQUID is essential to achieve a high multiplexing factor in the FDM system proposed for {\it Athena} \cite{denHartog14}.
 
\begin{figure}[htbp]
\centering
\includegraphics[width=.4\textwidth,keepaspectratio,
  angle=270]{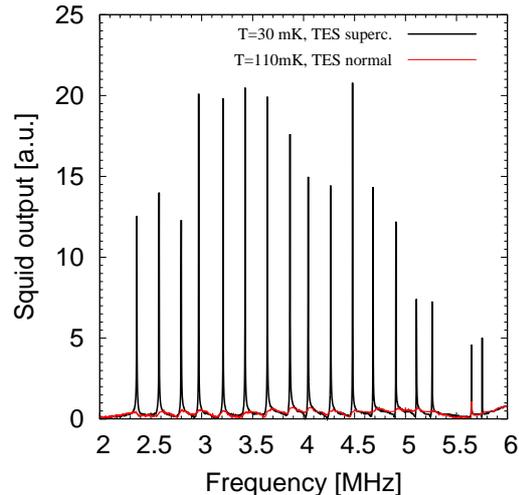}
\caption{ Frequency sweep via the ac-bias input port with TESs in superconducting ({\it black}) and normal ({\it red}) state. \label{sweepfreq}}
\end{figure}

In \figref{sweepfreq} we show the SQUID response of a frequency
sweep across the resonators via the ac-bias with the TESs in superconducting and in the normal state respectively.



\section{CONCLUSION} 

We fabricated and tested an FDM set-up for the single pixel
characterisation of TES based microcalorimeters under ac
bias in the  MHz frequency range. The set-up is designed to read out
18 pixels under ac bias and 2 pixels under AC or DC bias.

We validate the performance of the ac bias read-out  by characterising a the novel VTT
two-stage SQUID amplifier  and the lithographic high-$Q$ $LC$ filters at $T=25\,\mathrm{mK}$ in a removable probe of a He-free dilution refrigerator from Leiden Cryogenics. 
The two-stage SQUID amplifier developed at VTT has a measured flux noise at $25\, \mathrm{mK}$ equal to $1.8\cdot10^{-7}\,\Phi_0/Hz^{1/2}$ and an input current noise of   $1.2\cdot10^{-12}\,A/Hz^{1/2}$ , which corresponds to a coupled
energy resolution of $\epsilon_{ii}=20\,\hbar$ over the whole interesting frequency range from 2 to 5 MHz. 
  
We  have shown the progress in the fabrication of TES microcalorimeter array based on TiAu TESs and  large CuBi absorber. The pixels have been designed to meet the requirements of the X-IFU instrument on board of Athena.


 


\bibliography{GottardiSPIE2014}   
\bibliographystyle{spiebib}   

\end{document}